\documentclass[aps,twocolumn,showpacs,floats,prl]{revtex4}
\usepackage{graphicx}
\usepackage{dcolumn}
\usepackage{bm}
\usepackage{color}
\usepackage{amsmath}
\usepackage{subfigure}
\usepackage{enumitem}

\begin{document}

\author{L. Parisi}
\author{S. Giorgini}
\affiliation{Dipartimento di Fisica, Universit\`a di Trento and CNR-INO BEC Center, I-38123 Povo, Trento, Italy}

\title{Quantum Monte-Carlo study of the Bose polaron problem \\ in a one-dimensional gas with contact interactions} 

\begin{abstract} 
We present a theoretical study based upon quantum Monte Carlo methods of the Bose polaron in one-dimensional systems with contact interactions. In this instance of the problem of a single impurity immersed in a quantum bath, the medium is a Lieb-Liniger gas of bosons ranging from the weakly interacting to the Tonks-Girardeau regime, whereas the impurity is coupled to the bath via a different contact potential producing both repulsive and attractive interactions. Both the case of a mobile impurity, having the same mass as the particles in the medium, and of a static impurity with infinite mass are considered. We make use of exact numerical techniques that allow us to calculate the ground-state energy of the impurity, its effective mass as well as the contact parameter between the impurity and the bath. These quantities are investigated as a function of the strength of interactions between the impurity and the bath and within the bath. In particular, we find that the effective mass rapidly increases to very large values when the impurity gets strongly coupled to an otherwise weakly repulsive bath. This heavy impurity hardly moves within the medium, thereby realizing the ``self-localization'' regime of the Landau-Pekar polaron. Furthermore, we compare our results with predictions of perturbation theory valid for weak interactions and with exact solutions available when the bosons in the medium behave as impenetrable particles.  
\end{abstract}

\maketitle

\section{I. Introduction}

In recent years the problem of an impurity coupled to a quantum bath has received a large attention in the field of ultracold atoms both theoretically and experimentally. In particular, the fermionic version of this problem, {\it i.e.} when the bath is a spin-polarized Fermi sea, is already a well studied topic with different interesting perspectives addressing the attractive and repulsive branch as well as low dimensions~\cite{Review14}. On the contrary, for the bosonic counterpart corresponding to an impurity immersed in a Bose condensed medium, only very recently there have been experimental studies focusing on the excitation energy and spectral response of the polaron quasiparticle~\cite{Aarhus, JILA}. Previous experiments investigated mainly collisional and dissipation processes involving the bath~\cite{Widera12, Catani12, Oberthaler13}. On the theoretical side, the Bose polaron problem has been already addressed in a series of studies uitilizing different tools such as T-matrix~\cite{Rath13} and perturbation~\cite{Christensen15} approaches, variational wavefunction~\cite{DasSarma14, Levinsen15} as well as quantum Monte Carlo (QMC) methods~\cite{Ardila15, Ardila16}. 

Low dimensions, and in particular one-dimensional (1D) configurations, enrich the Bose polaron problem with some peculiar features that are worth investigating. First of all the enhanced role of quantum fluctuations capable to destroy the off-diagonal long-range order responsible for Bose-Einstein condensation even at zero temperature. Secondly, the possibility to achieve strongly correlated regimes for the quantum bath, where the bosons approach the so called Tonks-Girardeau (TG) limit of fermion-like impenetrable particles~\cite{Stoferle04, Paredes04, Weiss05, Haller09}. The use of confinement induced resonances of the $s$-wave scattering amplitude represents for these 1D systems a powerful tool, allowing for a wide tunability of both the interactions within the bath and between the bath and the impurity~\cite{Olshani98, Haller09, Catani12}. For example, in Ref.~\cite{Catani12} a 1D mixture of K impurities in a gas of Rb atoms was realized and different values of the impurity-bath interaction, obtained by varying the magnetic field, were probed. The gas parameter in the bath can also be tuned by changing the density or the effective mass of the atoms by means of a lattice potential~\cite{Paredes04, Weiss05}.

In this work we investigate theoretically the Bose polaron problem when the quantum bath is modeled by a Lieb-Liniger gas with contact interactions~\cite{LiebLiniger} and the impurity is coupled to the bath via a different $\delta$-like potential that can be both attractive and repulsive. Different regimes of the surrounding medium are considered and analyzed as a function of the impurity-boson coupling strength: i) the bath is weakly interacting and well described by the mean-field theory, ii) interactions in the bath are relatively strong and beyond mean-field effects are important, and finally iii) the bath is in the TG regime and some analytical results are available thanks to the Bose-Fermi mapping~\cite{McGuire1, McGuire2}. We characterize the Bose polaron by calculating its binding energy, effective mass, and contact parameter. Furthermore, we consider both the case of a mobile impurity having the same mass of the particles in the medium and the case of a static impurity, corresponding to the limit of a much larger impurity to boson mass ratio. 

The results show that the polaron energy reaches a constant value for large repulsive impurity-boson interaction, whereas in the opposite limit of large attraction the impurity gets deeply bound to the bath. Unless the bosons are impenetrable particles, this binding energy is found to be much larger than the one of the dimer state, the solution to the two-body problem in vacuum. The tendency of bosons attracted by the impurity to form large clusters involving many particles is also evident from the density profile of the medium around the impurity. The size of these clusters, as well as their binding energy, increases with decreasing interaction strength within the bath, indicating an instability of the weakly interacting Lieb-Liniger gas towards collapse around an attractive impurity. 

The effective mass of the polaron moving in a weakly interacting medium exhibits a sharp increase as a function of the coupling, both on the repulsive and on the attractive side of the impurity-bath interaction. Already for a moderate coupling strength the impurity gets very heavy, thus realizing the so-called ``self-localization'' regime predicted by Landau and Pekar for polarons in crystals~\cite{LandauPekar}. This picture was also proposed for Bose polarons in three dimensions (3D) on the basis of a Fr\"ohlich-type model~\cite{Cucchietti06, Kalas06, Jaksch08, Tempere09} which, however, does not capture the relevant physics at strong coupling~\cite{Christensen15, Ardila15}. On approaching the TG limit, instead, the effective mass increases sizeably only for large values of the coupling on the repulsive side and saturates at twice the bare mass on the attractive side. 

It is worth pointing out that the results mentioned above are directly relevant for experiments on few impurities in a quantum bath where the binding energy can be measured using spectroscopic techniques~\cite{Schirotzek09, Kohstall12, Koschorreck12, Aarhus, JILA} and the effective mass from the study of collective modes in harmonic traps~\cite{Nascimbene09, Catani12}.

The paper is organized as follows: in Sec.~II we introduce the Hamiltonian of the system and we review or derive some analytical results that are obtained using perturbation theory in the limit of weak interactions or the Bose-Fermi mapping in the TG limit. In Sec.~III we briefly provide some details about the QMC techniques used to address this specific problem and about the physical observables calculated in the simulations. Sec.~IV contains a detailed discussion of the results and a comparison with the analytical predictions available for specific regimes of parameters. Finally, in Sec.~V we draw our conclusions and we outline possible future prospects of this work.

\section{II. General theory}

We consider the following Hamiltonian 
\begin{eqnarray}
H&=&-\frac{\hbar^2}{2m_B}\sum_{i=1}^N \frac{\partial^2}{\partial x_i^2}+\sum_{i<j}g\delta(x_i-x_j)
\nonumber\\
&-& \frac{\hbar^2}{2m_I}\frac{\partial^2}{\partial x_\alpha^2}+\sum_{i=1}^N\tilde{g}\delta(x_i-x_\alpha) \;,
\label{Hamiltonian}
\end{eqnarray}
describing a 1D system of $N$ identical bosons with mass $m_B$ interacting via a repulsive contact potential of strength $g>0$. A single impurity of mass $m_I$ is coupled to the particles of the bath via another contact potential characterized by the strength $\tilde{g}$. The coordinates of the bosons and the impurity are denoted respectively by $x_i$ ($i=1,\dots,N$) and $x_\alpha$. We introduce the dimensionless parameters 
\begin{equation}
\gamma=\frac{gm_B}{\hbar^2n} \;\;\;\;\;\; \eta=\frac{2\tilde{g}}{\hbar^2n}\frac{m_Bm_I}{m_B+m_I} \;,
\label{parameters}
\end{equation}
where $n=N/L$ is the density of the bath in the 1D box of size $L$. The first parameter gives the strength of interactions within the bath which can range from the weakly correlated mean-field regime ($\gamma\ll1$) to the strongly correlated TG regime ($\gamma\gg1$), where bosons are impenetrable and behave similarly to a gas of spinless fermions. The parameter $\eta$ is instead related to the coupling between the impurity and the bath and, contrary to $\gamma$, can be either positive or negative depending on the sign of $\tilde{g}$. Notice that for equal masses ($m_B=m_I$) and for equal coupling strengths ($\tilde{g}=g$) the equality $\gamma=\eta$ holds. 

The above Hamiltonian (\ref{Hamiltonian}) of the clean bath without the impurity was solved exactly by Lieb and Liniger~\cite{LiebLiniger} for a system in the thermodynamic limit and with periodic boundary conditions. The ground-state energy is found as a function of the parameter $\gamma$ in the form
\begin{equation}
E_0=N\frac{\hbar^2n^2\pi^2}{2m_B}\epsilon(\gamma) \;,
\label{LiebLiniger}
\end{equation}
where the dimensionless energy per particle $\epsilon(\gamma)$ is obtained by solving a pair of coupled integral equations~\cite{LiebLiniger}. The scale of energy in Eq.~(\ref{LiebLiniger}) is chosen as $\epsilon_F=\frac{\hbar^2k_F^2}{2m_B}$, with $k_F=\pi n$, corresponding to the Fermi energy of a gas of spinless fermions with the same mass and density. Some limits of $\epsilon(\gamma)$ can be derived analytically: if $\gamma\ll1$, one finds $\epsilon\simeq\frac{\gamma}{\pi^2}(1-\frac{4}{3\pi}\sqrt{\gamma})$; in the opposite limit, $\gamma\gg1$, one has instead $\epsilon\simeq\frac{1}{3}(1-\frac{4}{\gamma})$. As we already mentioned, the former result holds in the weak-coupling regime and includes the Bogoliubov term and the first beyond mean-field correction whereas, in the latter, the leading term corresponds to the energy of the equivalent non-interacting Fermi gas.  

The binding energy of the impurity is defined as the energy difference between the ground state of the system with and without the impurity. The former can be written as 
 \begin{equation}
\tilde{E}_0=\frac{\hbar^2n^2\pi^2}{2m_B}\left[N\epsilon(\gamma)+\mu(\gamma,\eta)\right] \;,
\label{groundstate}
\end{equation}
allowing one to express the energy difference $\tilde{E}_0-E_0$ in terms of the dimensionless function $\mu(\gamma,\eta)$ which yields the polaron energy in units of the scale $\epsilon_F$.

In the limit of small $\gamma$ and small $\eta$ one can determine the energy and the effective mass of the polaron by using the perturbation approach based on the Bogolubov approximation as it is outlined in Ref.~\cite{Ardila15}. By introducing the mass ratio $w=\frac{m_B}{m_I}$ of the bosonic to the impurity mass, the result for the polaron energy reads:
\begin{eqnarray}
\mu(\gamma,\eta)&\simeq&\frac{\eta(1+w)}{\pi^2}\left(1- \frac{\eta}{\sqrt{8\gamma}}\frac{1+w}{\pi} \right.
\nonumber\\
&\times& \left. \int_0^\infty \frac{dx}{\sqrt{x+2}}\frac{1}{\sqrt{x^2+2x}+wx}\right) \;,
\label{enerpert}
\end{eqnarray}
and for the effective mass one finds
\begin{eqnarray}
\frac{m_I^\ast}{m_I}&\simeq&1+\frac{\eta^2}{\gamma^{3/2}}\frac{w(1+w)^2}{\sqrt{2}\pi}
\nonumber\\
&\times& \int_0^\infty \frac{dx}{\sqrt{x+2}}\frac{x}{(\sqrt{x^2+2x}+wx)^3} \;.
\label{meffpert}
\end{eqnarray}
The first term in Eq.~\ref{enerpert} is the mean-field contribution proportional to $\tilde{g}$, whereas next-to-leading order corrections to both $\mu$ and the effective mass are proportional to $\tilde{g}^2$ and are independent of the sign of the impurity-boson interaction.
We also notice that in the limit $\gamma\to0$, for a fixed value of $\eta$, both these corrections diverge, signalling an instability of the medium surrounding the impurity when boson-boson interactions are suppressed. The situation is different in the 3D case where, in the same limit, only the effective mass exhibits a divergent behavior~\cite{Ardila15}. 

We consider explicitly two particular values of the mass ratio $w$: the case of equal masses ($w=1$) and the case of a static impurity with infinite mass ($w=0$). For the former case, Eqs.~(\ref{enerpert}) and (\ref{meffpert}) give
\begin{eqnarray}
\mu(\gamma,\eta)&\simeq&\frac{2\eta}{\pi^2}\left(1-\frac{\eta}{\pi\sqrt{\gamma}}\right) \;,
\label{enerpert1}\\
\frac{m^\ast}{m}&\simeq&1+\frac{2}{3\pi}\frac{\eta^2}{\gamma^{3/2}} \;,
\label{meffpert1}
\end{eqnarray}
where we have set $m_B=m_I=m$. The binding energy of a static impurity, instead, is found to be:
\begin{equation}
\mu(\gamma,\eta)\simeq\frac{\eta}{\pi^2}\left(1-\frac{\eta}{4\sqrt{\gamma}}\right) \;.
\label{enerpert0}
\end{equation}
These two cases are of special interest because an exact solution is available when the bath is in the TG limit ($\gamma=\infty$). The binding energy and the effective mass of the equal mass case ($w=1$) were calculated by McGuire~\cite{McGuire1,McGuire2} with the result
\begin{equation}
\mu=\frac{2}{\pi}\left[\frac{\eta}{2\pi}+\arctan\frac{\eta}{2\pi}-\frac{\eta^2}{4\pi^2}\left(\frac{\pi}{2}-\arctan\frac{\eta}{2\pi} \right)\right]
\label{McGuire1}
\end{equation}
and the two results 
\begin{eqnarray}
\frac{m^\ast}{m}&=&\frac{2}{\pi}\frac{\left(\arctan\frac{2\pi}{\eta}\right)^2}{\arctan\frac{2\pi}{\eta}-\frac{2\pi/\eta}{1+\frac{4\pi^2}{\eta^2}}} \;,
\label{McGuire2-1}\\
\frac{m^\ast}{m}&=&\frac{2\left(1-\frac{1}{\pi}\arctan\frac{2\pi}{|\eta|}\right)^2} {1-\frac{1}{\pi}
\left(\arctan\frac{2\pi}{|\eta|}-\frac{2\pi/|\eta|}{1+\frac{4\pi^2}{\eta^2}} \right)} \;,
\label{McGuire2-2}
\end{eqnarray}
holding, respectively, for positive and negative values of the coupling constant $\tilde{g}$. Some limits of the above Eqs.~(\ref{McGuire1})-(\ref{McGuire2-2}) are worth discussing: i) If $\eta\to+\infty$ one finds $\mu\simeq1$, yielding a polaron energy equal to the chemical potential of the surrounding Fermi gas, whereas the effective mass diverges as $\frac{m^\ast}{m}\simeq\frac{3\eta}{2\pi^2}$. ii) If $\eta\to-\infty$, then $\mu\simeq-\frac{\eta^2}{2\pi^2}$ and $\frac{m^\ast}{m}\simeq2$, corresponding to the binding energy and the mass of a dimer in vacuum. 

In the case of a TG gas with a static impurity ($w=0$) one proceeds by considering the impurity in the center of a large box of size $L$ with impenetrable walls and by calculating the phase shift of each single-particle state generated by the impurity contact potential with strength $\tilde{g}$. The ground-state energy difference $\tilde{E}_0-E_0$ is readily calculated yielding the result
\begin{eqnarray}
\mu&=&\frac{1}{\pi}\left[\left(1+\frac{\eta^2}{4\pi^2}\right)\arctan\frac{\eta}{2\pi}+\frac{\eta}{2\pi}-\frac{\eta|\eta|}{8\pi}\right]
\nonumber\\
&-&[1-\theta(\eta)]\frac{\eta^2}{4\pi^2} \;.
\label{mustatic}
\end{eqnarray} 
The term involving the Heaviside function $\theta(x)$, where $\theta(x)=1$ if $x>0$ and zero otherwise, accounts for the binding energy of the dimer when $\tilde{g}<0$. Also in this case we can easily extract the following limiting behaviors: i) if $\eta\to+\infty$, then $\mu\simeq\frac{1}{2}$ and ii) if $\eta\to-\infty$ one finds the result $\mu\simeq-\frac{\eta^2}{4\pi^2}-\frac{1}{2}$. 
Notice that the energy of the mobile impurity (\ref{McGuire1}) in the limit of infinite repulsion is twice the corresponding energy of the static impurity (\ref{mustatic}), even though the effective mass (\ref{McGuire2-1}) diverges in the same limit. This is due to the kinetic energy contribution of the mobile impurity within the region of space delimited by the two nearest neighbour particles of the bath acting as impenetrable barriers.   
 
Another asymptotically exact result for a static impurity ($w=0$) is obtained when interactions within the bath are weak ($\gamma\ll1$) and there is a strong impurity-boson repulsion ($\eta\to+\infty$). In this case, the polaron energy coincides with the excitation energy of a dark soliton~\cite{Stringari-book}
\begin{equation}
\mu=\frac{8}{3\pi^2}\sqrt{\gamma} \;.
\label{soliton}
\end{equation}
This excited state of the gas is indeed stationary and is characterized by a zero in the density profile.

An important quantity describing the interaction between the impurity and the bath is the contact $C$, defined as the value of the boson density at the impurity position normalized by the bulk density: $C=\frac{n(x_\alpha)}{n}$. By using the Hellmann-Feynman theorem~\cite{Gangardt03} one can relate the value of $C$ to the derivative of the equation of state with respect to the impurity-boson coupling constant
\begin{equation}
C=\frac{d\tilde{E}_0}{nd\tilde{g}}=\frac{\pi^2m_I}{m_B+m_I}\frac{d}{d\eta}\mu(\gamma,\eta) \;.
\label{contact}
\end{equation}
In the case of equal masses ($w=1$) the contact parameter can be derived analytically in the weak-coupling limit, where Eq.~(\ref{enerpert1}) yields $C=1-\frac{2\eta}{\pi\sqrt{\gamma}}$, and in the TG limit, where from Eq.~(\ref{McGuire1}) one finds
\begin{equation}
C=1-\frac{\eta}{4}+\frac{\eta}{2\pi}\arctan\frac{\eta}{2\pi} \;.
\label{contactTG}
\end{equation}
From the above result we see that $C\simeq\frac{4\pi^2}{3\eta^2}$ when $\eta$ is large and positive and $C\simeq\frac{|\eta|}{2}$ in the opposite limit of large and negative values of $\eta$.

\section{III. Numerical method}

The calculations are performed using QMC numerical techniques. A thorough description of these methods can be found elsewhere~\cite{Reynolds-book}. Here we provide a detailed explanation of the choice of the wave function used as a trial many-body state in the variational Monte Carlo (VMC) and for importance sampling in the diffusion Monte Carlo (DMC) scheme. A similar scheme was used in the recent study of an impurity immersed in a 3D Bose-Einstein condensate~\cite{Ardila15}.

We consider a system of $N$ identical particles and one impurity contained in a 1D box of size $L$ with periodic boundary conditions. The trial wave function is comprised of two terms: the first takes into account correlations among the particles in the bath and the second between these particles and the impurity. Both pair correlation functions satisfy boundary conditions when the two particles meet to enforce the contact interactions present in the Hamiltonian (\ref{Hamiltonian}). The trial wave function is written as
\begin{equation}
\psi_T({\bf X})=\prod_{i<j}f_B(x_i-x_j)\prod_{i=1}^Nf_I(x_i-x_\alpha) \;,
\label{trial}
\end{equation}     
where the multidimensional vector ${\bf X}=(x_1,\dots,x_N,x_\alpha)$ denotes the coordinates of both the $N$ particles of the bath and the impurity. The boson-boson correlation factor is chosen as the solution of the two-body problem
\begin{equation}
f_B(x)=\cos\left(k|x|+\varphi_B(k)\right) \;, 
\label{fboson}
\end{equation}
if $|x|<X_B$ and $f_B(x)=1$ otherwise. The value of $k$ is fixed by the condition $kX_B+\varphi_B(k)=0$ ensuring that both $f_B$ and its first derivative are continuous at the matching point $X_B$. The phase shift results from the Bethe-Peierls contact condition imposed by the interatomic potential and is given by  $\varphi_B(k)=-\arctan\frac{gm_B}{2k\hbar^2}$. 
For the impurity-boson correlation term one should distinguish between the case of repulsive ($\tilde{g}>0$) and attractive ($\tilde{g}<0$) interactions. The former case can be dealt with in a way similar to the boson-boson interaction and one is lead to write 
\begin{equation}
f_I(x)=\cos\left(k|x|+\varphi_I(k)\right) \;,
\label{fimpurity}
\end{equation}
when $|x|<X_I$, while $f_I(x)=1$ if $|x|>X_I$. Here $k$ is chosen such that $kX_I+\varphi_I(k)=0$ in terms of the matching point $X_I$ and the phase shift is given by $\varphi_I(k)=-\arctan\frac{\tilde{g}m_Bm_I}{(m_B+m_I)k\hbar^2}$. In the case of attractive interactions a two-body bound state exists between the impurity and a particle of the bath and we choose the correlation term as a linear combination of the solutions corresponding to a negative two-body energy
\begin{equation}
f_I(x)=\frac{1}{2}\left(e^{-k(|x|-X_I)}+e^{k(|x|-X_I)}\right) \;,
\label{fimpurity-bis}
\end{equation}
when $|x|<X_I$ and $f_I(x)=1$ otherwise. The value of $k$ depends on the matching point $X_I$ and is determined through the contact condition imposed by the $\delta$-potential of strength $\tilde{g}$: $kX_I-\tanh^{-1}\frac{|\tilde{g}|m_Bm_I}{(m_B+m_I)k\hbar^2}=0$. The two matching points $X_B$ and $X_I$, entering respectively the boson-boson and the impurity-boson correlation term, are variational parameters optimized by minimizing the expectation value of the Hamiltonian on the many-body state $\psi_T$. 

Calculations are carried out by separately computing the ground-state energy of the system with $N$ bosons and one impurity, $E_0(N,1)$, and of the clean system with $N$ bosons 
only, $E_0(N)$. The polaron energy is determined as the difference $\mu=\frac{E_0(N,1)-E_0(N)}{\epsilon_F}$, in units of the Fermi energy. The effective mass is calculated by following the evolution in imaginary time of the impurity $\Delta x_\alpha(\tau)=x_\alpha(\tau)-x_\alpha(0)$, being $\tau=it/\hbar$, and by extracting the diffusion constant
\begin{equation}
\frac{m}{m^\ast}=\lim_{\tau\to\infty}\frac{\langle|\Delta x_\alpha(\tau)|^2\rangle}{2D\tau} \;,
\label{effmass}
\end{equation}
where $D=\frac{\hbar^2}{2m}$ is the diffusion constant of a free particle~\cite{Note1}. The total atom numbers $N$ range from $N=10$ to $N=200$ in order to control finite-size effects and extrapolate the results to the thermodynamic limit. 

The density profiles $n(x)$ of the bosons in the vicinity of the impurity are calculated by making use of the standard extrapolation technique~\cite{Mitas11} 
\begin{equation}
n(x)=2n_{\text{DMC}}(x)-n_{\text{VMC}}(x) \;,
\label{density}
\end{equation}
where $n_{\text{DMC}}$ is the ``mixed'' estimator of the density, obtained as the output of a DMC simulation, and $n_{\text{VMC}}$ is the ``variational'' estimator obtained from the corresponding VMC calculation. The contact parameter $C$ is extracted directly from the value of $n(x)$ at the position of the impurity.

\begin{figure*}
\scalebox{1.0}{\includegraphics[width=0.4\textwidth,height=6.5cm]{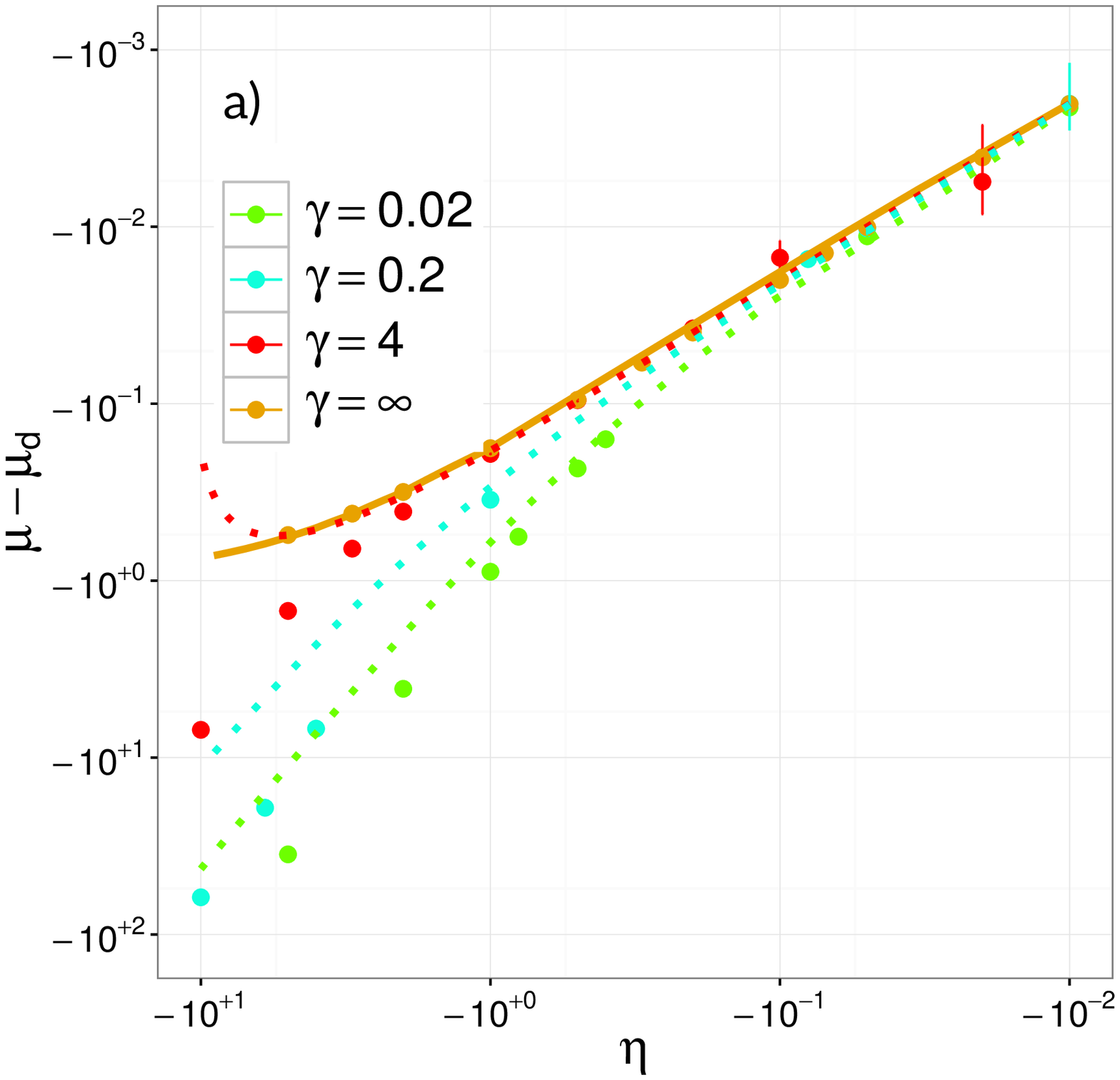}\includegraphics[width=0.4\textwidth,height=6.5cm]{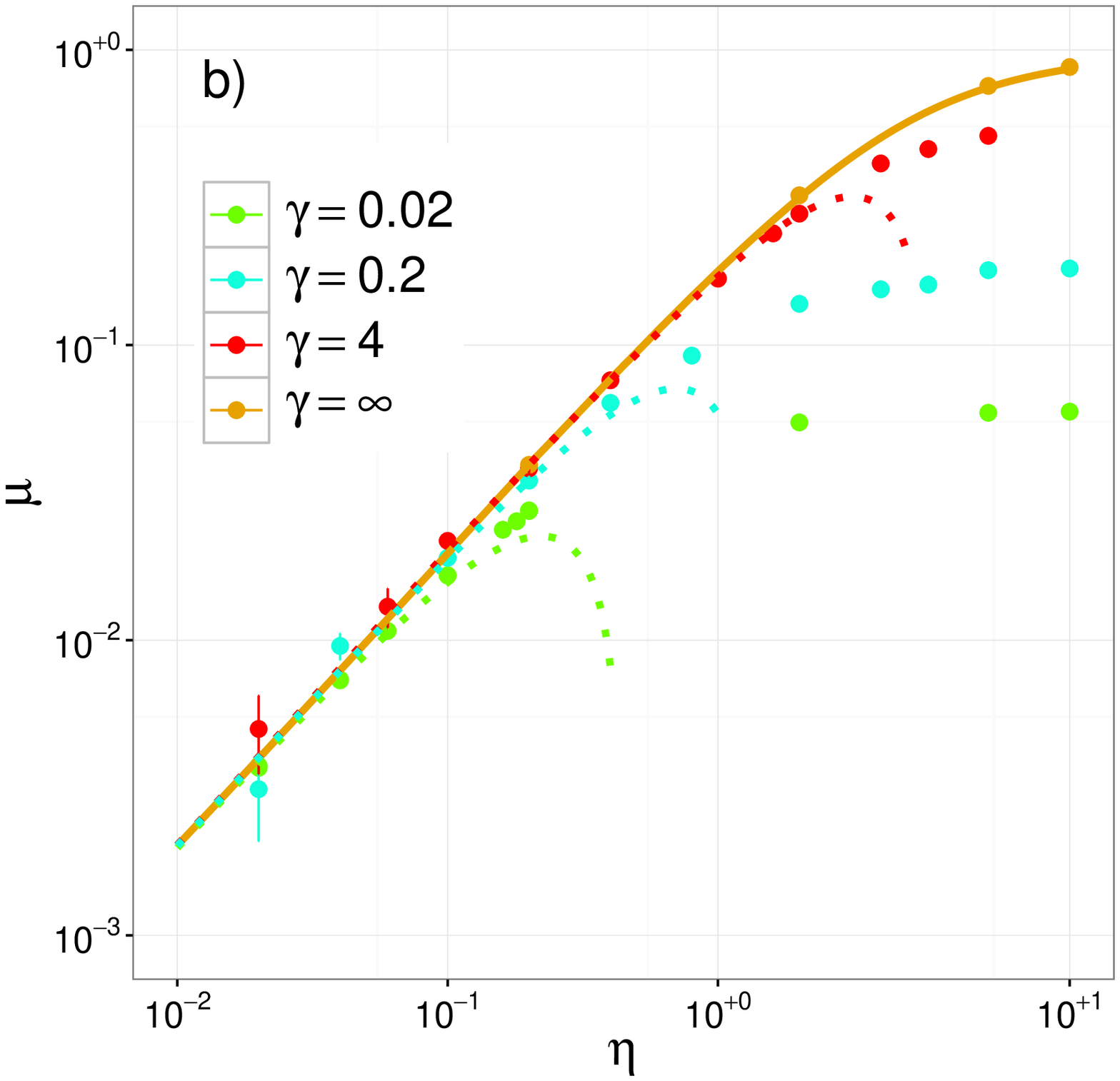}}
\caption{(Color online) Binding energy of the mobile impurity with mass ratio $w=1$ as a function of the impurity-boson interaction parameter $\eta$ and for different values of the coupling strength $\gamma$ within the bath. Panel (a) and (b) refer respectively to negative and positive values of $\tilde{g}$ and in panel (a) we subtracted from $\mu$ the contribution $\mu_d=-\frac{\eta^2}{2\pi^2}$ from the dimer bound state. The $\gamma=\infty$ curve corresponds to the exact result by McGuire [Eq.~(\ref{McGuire1})] in the TG limit. Dotted lines refer instead to the result of perturbation theory given in Eq.~(\ref{enerpert1}).}
\label{fig1}
\end{figure*}

\begin{figure*}
\scalebox{.95}{\includegraphics[width=0.4\textwidth,height=6.5cm]{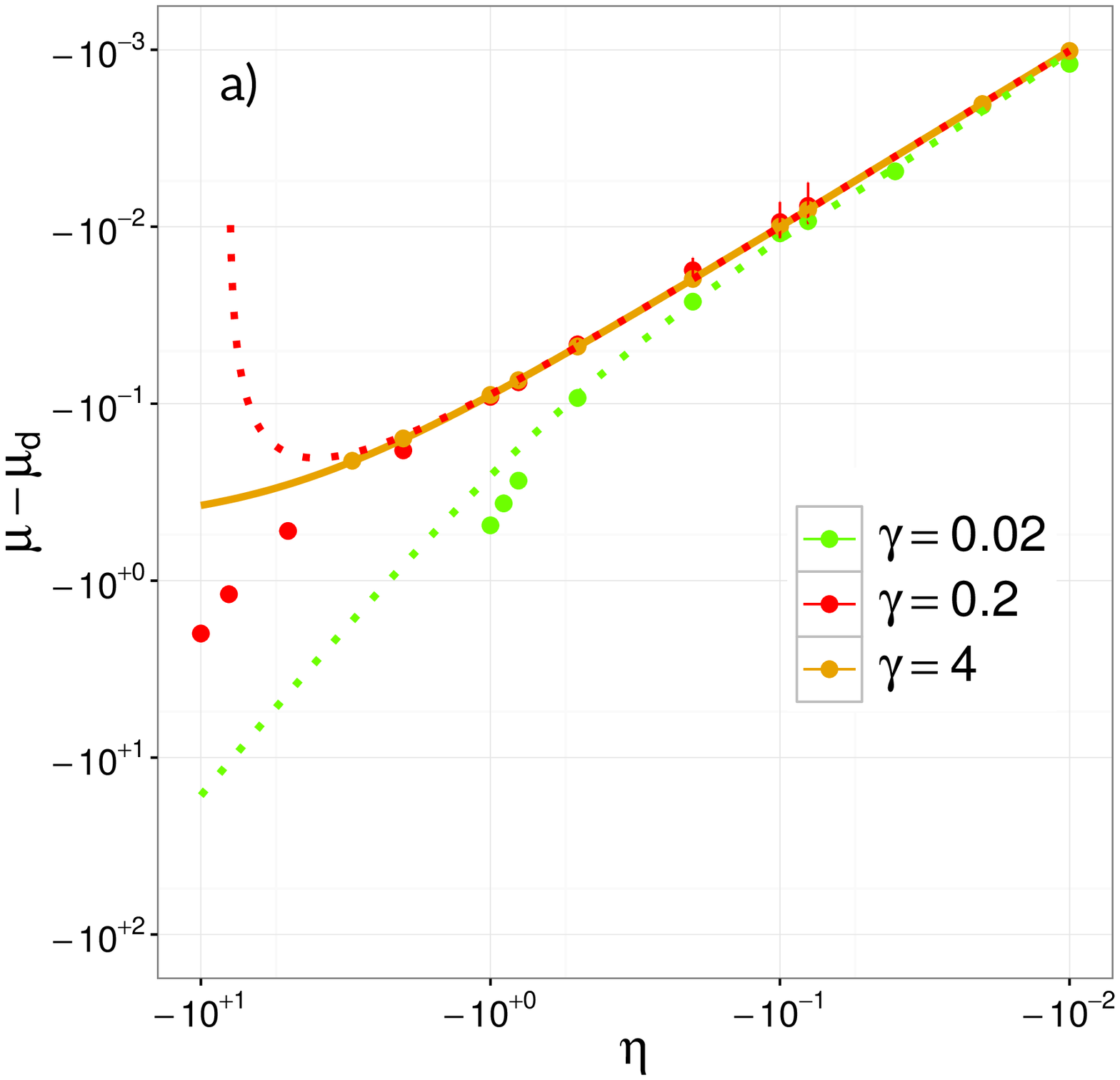}\includegraphics[width=0.4\textwidth,height=6.5cm]{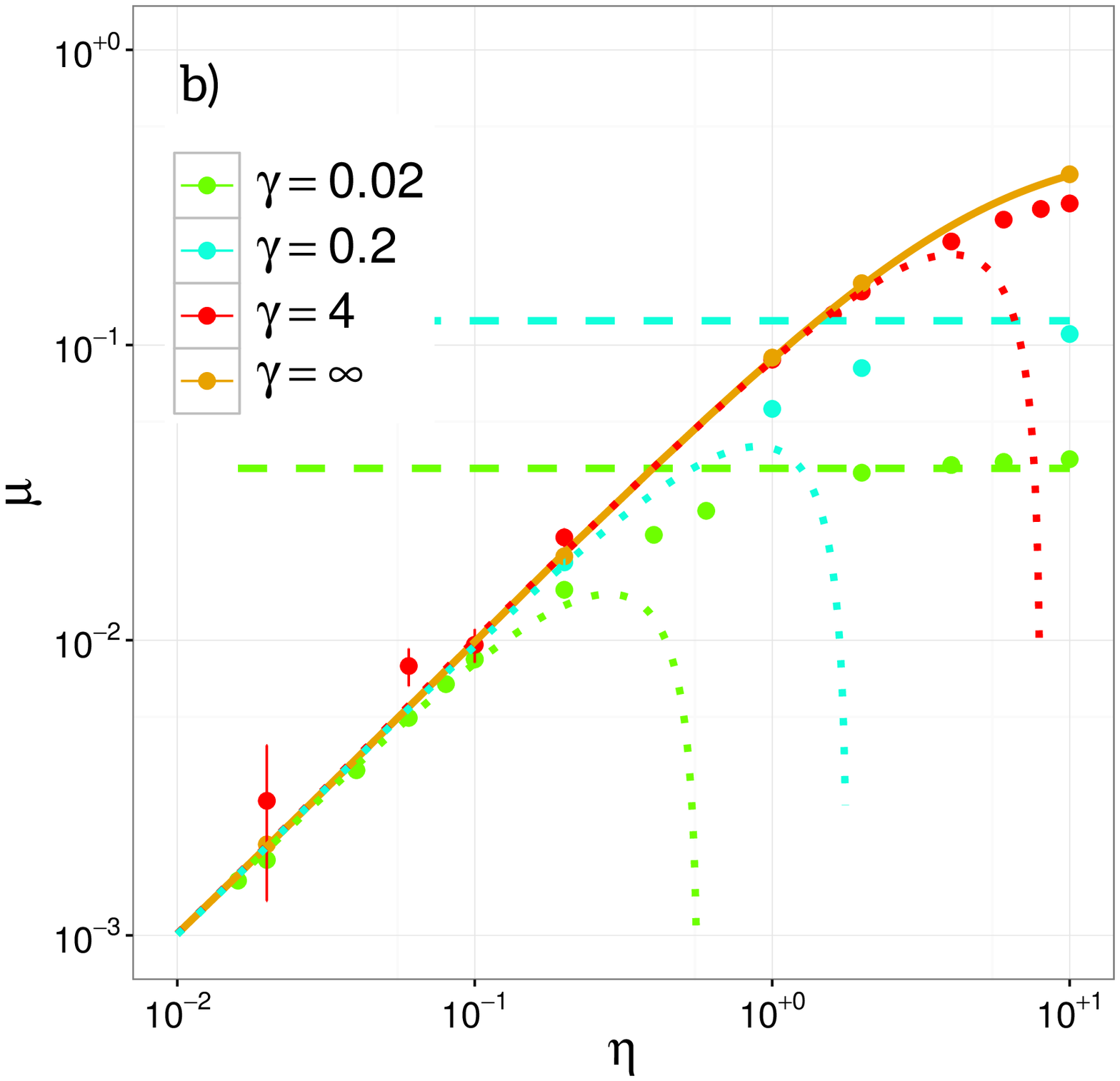}}
\caption{(Color online) Binding energy of the static impurity with mass ratio $w=0$ as a function of the impurity-boson interaction parameter $\eta$ and for different values of the coupling strength $\gamma$ within the bath. Panel (a) and (b) refer respectively to negative and positive values of $\tilde{g}$ and in panel (a) we subtracted from $\mu$ the contribution $\mu_d=-\frac{\eta^2}{4\pi^2}$ from the dimer bound state. The $\gamma=\infty$ curve corresponds to the exact result of Eq.~(\ref{mustatic}) in the TG limit. Dotted lines refer to the result of perturbation theory given in Eq.~(\ref{enerpert0}) and the dashed lines to the dark-soliton excitation energy (\ref{soliton}).}
\label{fig2}
\end{figure*}

\section{IV. Results}

We first discuss the results on the binding energy for the mobile ($w=1$) and the static ($w=0$) impurity. In Fig.~\ref{fig1} we show the energy of the mobile impurity as a function of the coupling strength $\eta$, ranging from large negative to large positive values, when the interaction parameter $\gamma$ in the bath is kept fixed. For this latter we consider values varying from the weakly coupled regime, $\gamma\ll1$, to the TG regime where $\gamma=\infty$. The corresponding results for the static impurity are presented in Fig~\ref{fig2}. Notice that for $\eta<0$ we subtract from the binding energy $\mu$ the contribution from the two-body bound state given by $\mu_d=-\eta^2\frac{1+w}{4\pi^2}$.

The exact polaron energies corresponding to the TG regime and given by Eqs.~(\ref{McGuire1}) and (\ref{mustatic}) are shown as solid lines in Fig.~\ref{fig1} and Fig.~\ref{fig2}, respectively for $w=1$ and for $w=0$. As an important benchmark test we find that the values of $\mu$ obtained from our QMC simulations with $\gamma=\infty$ perfectly reproduce these results. In the regime of small values of $|\eta|$, our results also recover the expansions from perturbation theory given by Eqs.~(\ref{enerpert1}) and (\ref{enerpert0}). In particular, both for $w=1$ and for $w=0$, we find that the range of values of $|\eta|$ where the perturbation expansion agrees well with the calculated polaron energy becomes larger as $\gamma$ increases. In fact we notice that, when $\gamma$ is large, from Eq.~(\ref{enerpert}) one finds $\mu\simeq\frac{\eta(1+w)}{\pi^2}$. Remarkably this result agrees with the expansion of Eq.~(\ref{McGuire1}) and (\ref{mustatic}), respectively for the mobile and static impurity, up to values of the impurity-boson coupling constant on the order of $|\eta|\simeq1$. For smaller values of $\gamma$, typically $\gamma<1$, the applicability of perturbation theory is instead limited to the region where $|\eta|\lesssim\sqrt{\gamma}$. 

If $\eta$ is large and positive the energy of both the mobile and the static impurity tends to saturate to a value that becomes smaller with decreasing $\gamma$ [see panel (b) of Fig.~\ref{fig1} and Fig.~\ref{fig2}]. In the TG regime ($\gamma=\infty$) this asymptotic energy coincides with the energy $\epsilon_F$ of adding an extra particle to the bath in the mobile case and with $\epsilon_F/2$ in the static case. As already discussed in Sec.~II, this energy difference arises from the kinetic energy contribution of the mobile impurity. A similar difference persists also for smaller values of $\gamma$: for example, at $\gamma=0.02$ and $\eta=10$, we find $\mu=0.059(1)$ and $\mu=0.041(1)$ respectively for the $w=1$ and $w=0$ case. We notice that, in this latter limit of large $\eta$ and small $\gamma$, the energy of the static impurity is expected to coincide with the excitation energy (\ref{soliton}) of a dark soliton as determined using the mean-field Gross-Pitaevskii equation. Indeed, in panel b) of Fig.~\ref{fig2}, good agreement between the two energies is found for $\gamma=0.02$ and also for $\gamma=0.2$. 

In the opposite regime of large and negative values of $\eta$, the energy difference $\mu-\mu_d$ tends, when $\gamma=\infty$, respectively to $-1$ and to $-1/2$ in the mobile and in the static case. Here, the impurity forms a two-body bound state with one of the particles of the medium which is then missing from the Fermi sea of the TG gas. Notice that, similarly to the case of $\eta>0$, the binding energy $\mu$ of the mobile impurity is larger by a factor of two compared to the energy of the static one. For smaller values of $\gamma$ our results indicate that $\mu-\mu_d$ is always negative and grows unbounded as $\eta\to-\infty$ [see panel (a) of Fig.~\ref{fig1} and Fig.~\ref{fig2}]. This behavior arises because, for attractive interactions between the impurity and the bath and for not too strong repulsion within the bath, many particles of the medium tend to cluster around the impurity producing a large negative binding energy for the polaron.

An interesting question concerns the value of the polaron energy when $\eta$ is fixed and the interaction strength within the bath gets weaker and weaker ($\gamma\to0$). At the level of perturbation theory [see Eq.~(\ref{enerpert})] the answer to this question is that $\mu$ becomes large and negative irrespective of the sign of $\eta$. This result differs from the corresponding situation in 3D where perturbation theory predicts that, when interactions within the bath vanish, the polaron energy reduces to the mean-field value, proportional to the interspecies coupling constant~\cite{Ardila15}. For positive values of $\eta$, a proper answer to the question, going beyond the result of perturbation theory, is provided by panel (b) of Fig.~\ref{fig1} and Fig.~\ref{fig2}: when $\gamma\ll\eta$ the polaron energy decreases to zero and for the static impurity the behavior of $\mu$ is correctly described by the dark-soliton excitation energy given in Eq.~(\ref{soliton}). 

\begin{figure}
\begin{center}
\includegraphics[width=7.0cm]{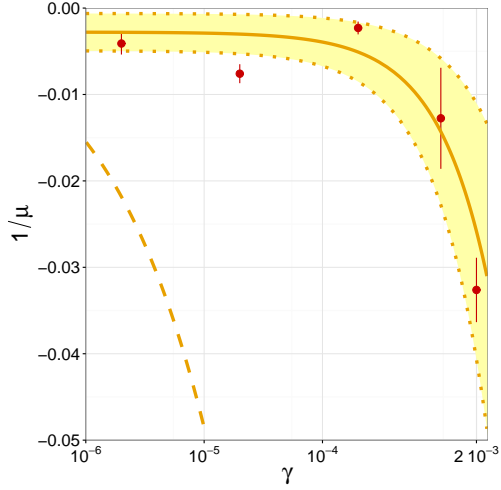}
\caption{(Color online) Inverse binding energy of the mobile impurity with $w=1$ as a function of the parameter $\gamma$ and for the fixed value $\eta=-1$ of the impurity-boson coupling constant. The line is a linear fit to the data and the shadow region shows the statistical uncertainty of the fit. The dashed line is the prediction from perturbation theory given in Eq.~(\ref{enerpert1}).}
\label{fig3}
\end{center}
\end{figure}

In the case of $\eta<0$,  panel (a) of Fig.~\ref{fig1} and Fig.~\ref{fig2} shows that the binding energy of the impurity grows large and negative as $\gamma$ decreases for a fixed value of $\eta$. However, in this limit, one can expect that either i) $\mu\to-\infty$, indicating the instability of the non-interacting bath in the presence of the impurity, or ii) $\mu$ saturates to a finite energy, indicating that even a tiny repulsion in the bath is enough to stabilize the polaron. We notice that the value of $\mu(\gamma,\eta)$ refers to the polaron energy in the thermodynamic limit and that the $\gamma\to0$ limit is intended to be taken after the one of $N\to\infty$. Of course, in the opposite case of a strictly non-interacting bath with a finite number of particles, the polaron energy would trivially diverge when the number $N$ increases. We address the question for the mobile impurity with mass ratio $w=1$, when the interaction between the impurity and the bath is attractive and kept fixed at the value $\eta=-1$. The results of the inverse energy $1/\mu$ are shown in Fig.~\ref{fig3} as a function of decreasing values of $\gamma$. We find that $1/\mu$ decreases in absolute value as $\gamma$ decreases, even though the result (\ref{enerpert1}) of perturbation theory fails completely in describing the trend of the calculated binding energies. A simple linear fit to the data extrapolates to a value compatible with $1/\mu=0$ when $\gamma=0$, given that error bars are significantly large. Our findings are thus compatible with the above case i), showing an instability of the weakly repulsive bath towards a collapse around the impurity. A similar behavior is expected for the static impurity with $w=0$. This latter result is in contrast with the binding energy of a static impurity in 3D and resonantly interacting with the medium which was found to approach a finite value in the limit of a vanishing repulsion within the bath~\cite{Ardila16}.

\begin{figure}
\begin{center}
\includegraphics[width=7.0cm]{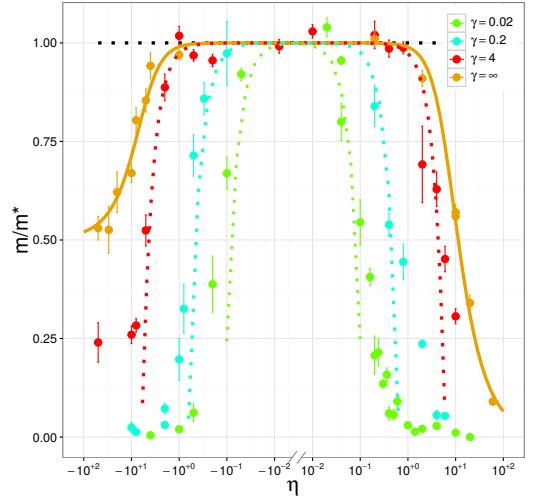}
\caption{(Color online) Inverse effective mass of the mobile impurity with mass ratio $w=1$ as a function of the impurity-boson interaction parameter $\eta$ and for different values of the coupling strength $\gamma$ within the bath. Both values corresponding to $\eta$ positive and negative are shown in the same graph. The $\gamma=\infty$ curve corresponds to the exact results by McGuire of Eqs.~(\ref{McGuire2-1}) and (\ref{McGuire2-2}) in the TG limit. Dotted lines refer to the result of perturbation theory given in Eq.~(\ref{meffpert1}).}
\label{fig4}
\end{center}
\end{figure}

In Fig.~\ref{fig4} we show the results of the polaron effective mass in the case of the mobile impurity with mass ratio $w=1$. We find that, for a given value of $\gamma$, the inverse effective mass decreases as $|\eta|$ increases both for repulsive and attractive interactions. We also notice that, in the TG limit of $\gamma=\infty$, we recover the exact results obtained by McGuire and given in Eqs.~(\ref{McGuire2-1}) and (\ref{McGuire2-2}). Furthermore, the comparison with the prediction (\ref{meffpert}) of perturbation theory shows that, similarly to the case of the energy $\mu$, the range of values of $|\eta|$ where agreement is found gets larger as $\gamma$ increases. 

On the attractive side of impurity-bath interactions, the exact TG-gas result in Eq.~(\ref{McGuire2-2}) yields $m^\ast\to2m$ in the limit of $\eta$ large and negative. For the strongly interacting medium with $\gamma=4$ we find in the same limit that the effective mass seems to saturate to $m^\ast\simeq4m$. One should stress here that for large attractions the calculation of the effective mass requires increasingly longer simulation times resulting in larger error bars. For smaller values of $\gamma$, both on the attractive and on the repulsive side, the value of $m^\ast/m$ becomes very large for $|\eta|\gtrsim10$, if $\gamma=0.2$, and already for $|\eta|\gtrsim1$ if $\gamma=0.02$. It is worth noticing that this rapid increase of the effective mass as a function of the impurity-bath coupling does not occur in the 3D counterpart of the Bose polaron problem, where in the limit of resonant interaction between the impurity and the bath one finds $m^\ast/m\lesssim2$ (see Ref.~\cite{Ardila15}). 

An increase of the effective mass as a function of the impurity-bath coupling has been reported in the experiment of Ref.~\cite{Catani12} both for attractive and repulsive interactions. In this experiment a cloud of K impurities immersed in a 1D bath of Rb atoms is suddenly released after compression with an optical potential and the rate of increase of its axial size is measured for different fixed values of the interaction strength between the impurities and the bath. The connection with the effective mass of the impurities is provided by interpreting the normalized width of the cloud with $\sqrt{m_I/m_I^\ast}$. The coupling constant of the bath was $\gamma\simeq1$ and values of $\eta$ as large as $|\eta|\simeq10$ were produced, resulting in a maximum measured decrease of the normalized width by a factor of roughly 0.6.

\begin{figure}
\begin{center}
\includegraphics[width=8.0cm]{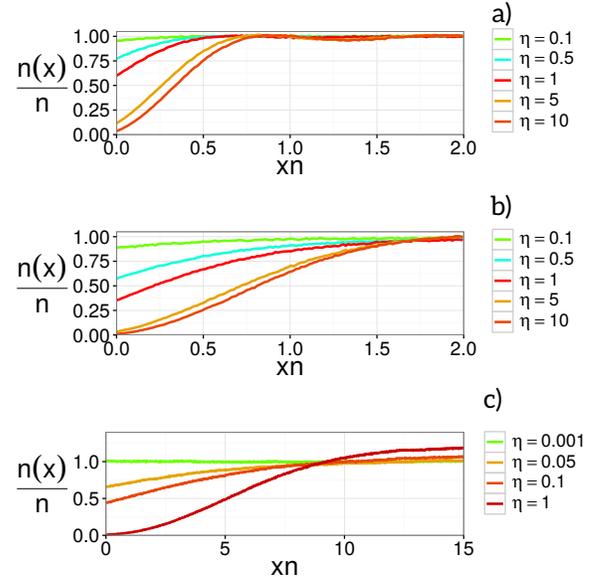}
\caption{(Color online) Density profile of the bath as a function of the distance from the mobile impurity with mass ratio $w=1$. Distances are in units of the inverse 1D density $1/n$. The different curves correspond to different values of the parameter $\eta$ characterizing the strength of impurity-bath repulsive interactions. The various panels refer to different values of the coupling constant within the bath: panel (a) $\gamma=\infty$, panel (b) $\gamma=2$ and panel (c) $\gamma=0.02$.}
\label{fig5}
\end{center}
\end{figure}

\begin{figure}
\begin{center}
\includegraphics[width=8.0cm]{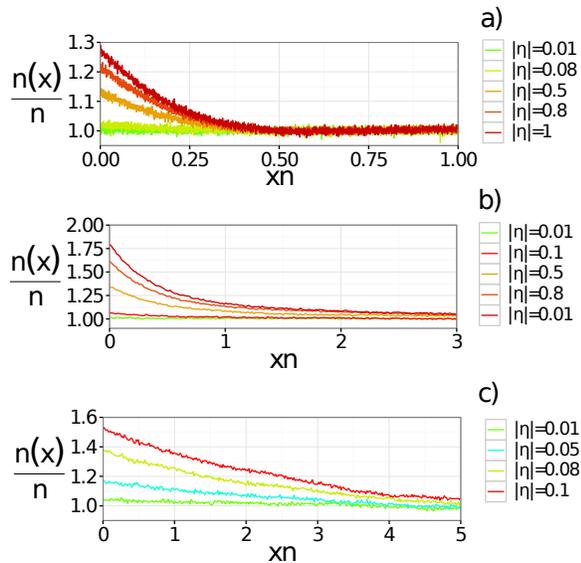}
\caption{(Color online) Density profile of the bath as a function of the distance from the mobile impurity with mass ratio $w=1$. Distances are in units of the inverse 1D density $1/n$. The different curves correspond to different values of the parameter $|\eta|$ characterizing the strength of impurity-bath attractive interactions. The various panels refer to different values of the coupling constant within the bath: panel (a) $\gamma=\infty$, panel (b) $\gamma=2$ and panel (c) $\gamma=0.02$.}
\label{fig6}
\end{center}
\end{figure}

In Fig.~\ref{fig5} and Fig.~\ref{fig6} we show the density profiles of the bath as a function of the distance from the mobile impurity with mass ratio $w=1$ and for different values of both the impurity-boson and the boson-boson coupling constant. Fig.~\ref{fig5} refers to repulsive interactions between the impurity and the bath, whereas Fig.~\ref{fig6} refers to attractive interactions. By increasing the value of $\eta$ in the case of repulsive interactions, the density of bosons in the close vicinity of the impurity decreases until a hole, completely empty of particles, is created for very large $\eta$. The size of the hole strongly depends on the interaction parameter within the bath: it is on the order of the interparticle distance for the largest value ($\gamma=\infty$) and it extends to up to $\sim10$ interparticle distances for the smallest one ($\gamma=0.02$) [see panels (a)-(c) in Fig.~\ref{fig5}]. We also notice that, for $\gamma=\infty$ and for the largest value of $\eta$, Friedel-type oscillations, typical of the fermionic nature of the TG gas, are visible in the density profile. 

The results of Fig.~\ref{fig6}, instead, feature a peak of the boson density around the position of the impurity which becomes higher as the strength of the attraction increases. Also in this case the size of the peak gets larger as $\gamma$ decreases. For the largest $\gamma$ [see panel (a) in Fig.~\ref{fig6}] the size of the peak is smaller than the interparticle distance and at large values of $|\eta|$ only one particle of the bath is, on average, close to the impurity forming a bound dimer with energy $\mu_d$. Indeed, as already mentioned when discussing panel (a) of Fig.~\ref{fig1}, the energy of a polaron in a TG gas tends to $\mu_d$ when $\eta$ is large and negative. On the contrary, as the value of $\gamma$ decreases, the density peak becomes wider and involves more and more particles of the bath [see panels (b) and (c) in Fig.~\ref{fig6}]. As a consequence, the binding energy of the impurity increases in absolute value and one is approaching the situation of an unstable weakly interacting gas as shown in Fig.~\ref{fig3}.

\begin{figure}
\begin{center}
\includegraphics[width=7.5cm]{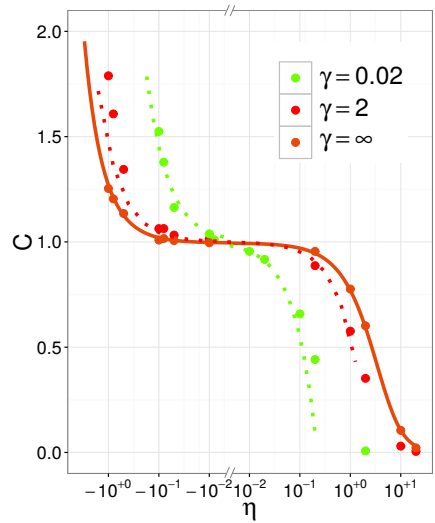}
\caption{(Color online) Contact parameter $C$ of a mobile impurity with mass ratio $w=1$ as a function of the impurity-boson interaction parameter $\eta$ and for different values of the coupling strength $\gamma$ within the bath. Both values corresponding to $\eta$ positive and negative are shown in the same graph. The solid line corresponds to the exact result (\ref{contactTG}) holding in the TG limit. Dotted lines refer to the perturbative result holding for small values of $\eta$.}
\label{fig7}
\end{center}
\end{figure}

Finally, in Fig~\ref{fig7}, we show the results on the contact parameter $C$ defined in Eq.~(\ref{contact}). As explained in the previous section, the value of $C$ is determined from the boson density $n(x)$ at the impurity position normalized by the bulk density $n$. We notice that we reproduce the exact result of Eq.~(\ref{contactTG}) in the TG regime. Furthermore, as compared to the $\gamma=\infty$ case, we see that for smaller values of $\gamma$ the contact parameter drops faster with increasing positive $\eta$ and diverges faster with increasing negative $\eta$.

Qualitatively similar results for the density profiles and the contact parameter are obtained in the case of the static impurity with mass ratio $w=0$.  

\section{V. Conclusions}

By using ``exact" QMC numerical methods we investigated the properties of a Bose polaron in 1D as a function of both the coupling strength between the impurity and the bath and within the bath. For a given impurity-bath interaction strength we find that the repulsive polaron can never exceed the energy reached when the bath is in the TG regime. On the contrary, the binding energy of the attractive polaron lies always below the energy of a dimer in vacuum and becomes increasingly large as the repulsion within the medium is reduced, thereby  signalling an instability of the weakly interacting gas towards collapse around the impurity position. Furthermore, in the regime of a weakly repulsive medium, the polaron effective mass is found to increase sharply with the strength of the impurity-bath coupling. Such a heavy impurity, practically immobile within the medium, realizes the long-sought after regime of ``self-localization" of the strongly coupled Landau-Pekar polaron. 

Interesting future prospects of this work include the study of a finite concentration of impurities in a 1D quantum bath. This topic naturally leads to the issue of interaction between impurities, phase separation in the two-component gas and the properties of magnetic excitations in the miscible mixture.  

\section*{Acknowledgements}

Useful discussions with G.E. Astrakharchik are gratefully acknowledged. This work was supported by the QUIC grant of the Horizon2020 FET program and by Provincia Autonoma di Trento.

\end{document}